\documentclass[english]{article}
\usepackage[T1]{fontenc}
\usepackage[latin9]{inputenc}
\usepackage{color}
\usepackage{textcomp}
\usepackage{amstext}
\usepackage{amssymb}
\usepackage{stackrel}
\usepackage{graphicx}
\usepackage{esint}
\usepackage{cite}
\usepackage{epstopdf}
\usepackage{epsfig}
\usepackage{subfigure,array}

\makeatletter

\usepackage[a4paper]{geometry}
\makeatother

\usepackage{babel}
\begin{document}

\title{Higher-order nonclassical properties of a shifted symmetric cat state
and a one-dimensional continuous superposition of coherent states}

\author{Nasir Alam$^{\dagger}$, Kathakali Mandal{\normalsize{}$^{\mathsection}$}
and Anirban Pathak$^{\dagger,}$\thanks{anirban.pathak@jiit.ac.in}\\$^{\dagger}$Jaypee Institute of Information Technology, A-10, Sector-62,
Noida, UP-201309, India\\
$^{\mathsection}$Jaypee Institute of Information Technology, Sector-128,
Noida, UP-201304, India}

\maketitle

\begin{abstract}
Role of quantum interference in the origin of higher-order nonclassical
characteristics of radiation field has been probed vis-\`{a}-vis a discrete
and a continuous superposition of coherent states. Specifically, the
possibilities of observing higher-order nonclassical properties (e.g.,
higher-order antibunching (HOA), higher-order sub-Poissonian photon
statistics (HOSPS), higher-order squeezing (HOS) of Hong-Mandel type
and Hillery type) have been investigated using a shifted symmetric
cat state that reduces to Yurke-Stoler, even and odd coherent states
at various limits. This shifted symmetric cat state which can be viewed
as a discrete superposition of coherent states is found to show HOA
and HOSPS. Similarly, higher-order nonclassical properties of a one-dimensional
continuous superposition of coherent states is also studied here.
The investigation has revealed the existence of HOS and HOSPS in the
one-dimensional continuous superposition of coherent states studied
here. Effect of non-Gaussianity inducing operations (e.g., photon
addition and addition followed by subtraction) on these superposition states
have also been investigated. Finally, some comparisons have been made between
the higher-order nonclassical properties of discrete and continuous superposition
of coherent states. 

\end{abstract}

\section{Introduction}

It is well known that being a linear theory, quantum mechanics allows
the existence of various types of superposition states. In fact, the
superposition principle is at the heart of quantum mechanics \cite{buvzek1995quantum} and this is what leads no quantum supremacy (\cite{harrow2017quantum} and references therein).
However, linearity is not a unique property of quantum mechanics.
Even in classical optics, we observe interference which is a manifestation
of the superposition principle through first order correlation. First
order correlation or first order interference are not strong
enough to reveal unique (nonclassical) properties of quantum mechanics
which have no classical analogue. To witness the actual beauty, power
and mystery of quantum mechanics, we need to study second or higher-order
correlations. In fact, there exists a necessary and sufficient criterion
that identifies nonclassical state. Specifically, negative values
of Glauber-Sudarshan $P$-function \cite{glauber1963coherent,sudarshan1963equivalence}
defines a quantum state as nonclssical state (i.e., as a state having
no classical analogue). Unfortunately, except a single proposal for the measurement
of $P$-function in a special case \cite{kiesel2008experimental},
there does not exist any method for experimental determination of
$P$-function. Consequently, several operational criterion for witnessing
nonclassicality have been developed {(}\cite{miranowicz2010testing}
and references therein{)}. Any finite set of these witnesses of nonclassicality
is only sufficient \cite{richter2002nonclassicality}. So, failure
of any of these criteria does not imply classicality of the state,
but the fulfillment  of a criterion implies nonclassicality.
These witnesses of nonclassicality can be expressed as the moments of annihilation and
creation operators, and a nonclassical property witnessed through
a moment based criterion that witnesses a nonclassical feature through
a second order correlation (involves terms up to $4$th order in annihilation
and creation operators) is called a lower-order nonclassical property.
Naturally, higher-order nonclassicality refers to the nonclassical features
witnessed via higher-order correlations. Most frequently studied higher-order
nonclassical features are higher-order antibunching (HOA) \cite{pathak2006control}, higher-order sub-Poissonian photon statistics (HOSPS) \cite{prakash2006higher,verma2010generalized},
higher-order squeezing (HOS) of Hillery type \cite{hillery1987amplitude} and Hong-Mandel type
\cite{hong1985higher}. All of these nonclassical features have lower-order counterparts which have been studied more rigorously \cite{pathak2018classical,dodonov2002nonclassical}.
However, in the recent past, much attention has been given to the higher-order
nonclassical states because of a good number of successful experimental
characterization of such states \cite{avenhaus2010accessing,allevi2012high,allevi2012measuring,pevrina2017higher}. 

The experimental success in detecting higher-order nonclassicality
and the fact that weak nonclassicalities not-detected by their lower-order counterparts can be detected by the higher-order nonclassicality
criteria \cite{allevi2012high,allevi2012measuring} have led to a large number of theoretical works, too. In
fact, HOA has been reported in optomechanical and optomechanical-like system \cite{alam2017lower}, finite dimensional coherent state \cite{alam2017higher}, optical coupler \cite {thapliyal2014higher}, most general form of the hyper Raman process \cite{thapliyal2017nonclassicality}, etc.,  
HOSPS has been reported in finite dimensional coherent state \cite{alam2017higher}, photon added and subtracted squeezed coherent states \cite{thapliyal2017comparison}, various intermediate states \cite{verma2010generalized}, etc., 
and HOS has been reported in optomechanical and optomechanical-like system \cite{alam2017lower}, finite dimensional coherent state \cite{alam2017higher} and a pair of anharmonic oscillators \cite{alam2016nonclassical}.
However, to the best of our knowledge, until now no effort has been
made to investigate higher-order nonclassical properties of (discrete
and continuous) superposition of coherent states. It has not even
been studied for the relatively simple superposition states (namely
cat and cat-type states) that can be produced and manipulated in the
lab. To be precise, we are interested in the higher-order nonclassical
properties of a shifted symmetric cat state which reduces to various well known
superposition states like, odd and even coherent states, Yurke-Stoler
coherent state \cite{yurke1986generating} in different limits. We
are also interested in the higher-order nonclassial properties of
a one-dimensional continuous superposition of coherent states. It may be noted that lower-order nonclassical properties of these states have been studied since
long (cf. \cite{buvzek1995quantum,ren2017non,zeng2007nonclassical}
and references therein).

Apart from what stated above, this paper
is also motivated by the fact that different types of cat states have
now been produced in labs using various technologies \cite{gao2010experimental,leibfried2005creation},
and their applications have recently been reported in various tasks,
especially in connection with the rapidly developing field of quantum
information processing \cite{nielsen2002quantum}. In addition, applications
of non-Gaussian states \cite{Agarwal2013quantum} and effect of non-Gaussianity-inducing
operations (e.g., photon addition, photon subtraction and photon addition followed by subtraction) on the nonclassical
properties \cite{thapliyal2017comparison} have been studied in detail
in the recent past. However, no attention has yet been given to the role
of non-Gaussianity-inducing operations on the higher-order nonclassicality
witnesses in general and on the higher-order nonclassical properties
of shifted symmetric cat state (and  its special
cases) and the continuous superposition of coherent states 
in particular. This work is also slightly motivated by the recent
work of Ren, Du and Yu \cite{ren2017non}, where they have shown that
the depths of lower-order nonclassicality witnesses can be controlled
by controlling the relative phase between the opposite coherent states
that are superposed to produce the shifted cat state.

Keeping the above in mind, in what follows, we investigate the possibilities
of observing HOA, HOSPS, HOS of Hillery  type and Hong-Mandel type in
the shifted symmetric cat state and a one-dimensional continuous superposition of
coherent states. We have studied the variation of these higher-order nonclassicality witnesses
with respect to a controllable relative phase and average photon number.
Similarly, variations of higher-order nonclassicality witnesses have
also been reported for three special cases- even coherent state, odd
coherent state and Yurke-Stoler coherent state. Interestingly, it
is observed that HOA and HOSPS witness is observed for odd coherent
state. Subsequently, effect of photon addition and photon addition
then subtraction on the witnesses of higher-order nonclassicalities have been investigated with a particular focus on HOS. It's observed that nonclassical features
not observed in the original state can be observed after application
of these non-Gaussianity inducing operations. For example, HOS is
not observed for the discrete superposition of the coherent states,
but after photon addition HOS can be observed. Finally, it is noted that
the method followed here can also be used to study a set of similar states
that are produced by suitable superposition of coherent states.

The rest of the paper is organized as follows. In the Sec. \ref{sec:States-of-interest:},
we have introduced the discrete and continuous
superposition of coherent states studied in this paper. In Sec. \ref{sec:Witnesses-of-higher},
we have reported the existence of the various higher-order
nonclassicalities in the superposition states described in the previous section. In Sec. \ref{sec:Effect-of-photon}, we have investigated
the effect of non-Gaussianity inducing operations on the witnesses of higher-order nonclassicality. Finally,
the article is concluded in Sec. \ref{sec:CONCLUSION}.

\section{States of interest \label{sec:States-of-interest:}}

As mentioned above, in this section, we would describe the superposition
states that we wish to study in this work. Specifically, we will focus
on a discrete superposition and a continuous superposition of coherent
states. To begin with, we will describe a particularly interesting
discrete superposition of coherent states- a shifted symmetric cat
state. We will also mention, its special cases. It will be followed
by a brief description of a one-dimensional continuous superposition
of coherent states.

\subsection{Discrete superposition of coherent state: Shifted symmetric cat state}

{Let us begin our discussion with the standard definition
of symmetric cat state $|S\rangle$ and asymmetric cat state $|A\rangle$,
which are conventionally defined as the superposition of two opposite}{
coherent states (which are $\pi$ out of phase with respect to each
other) as follows \cite{cochrane1999macroscopically}. 
\begin{equation}
\begin{array}{lcl}
|S\rangle & = & N_{+}\left[|\alpha\rangle+|-\alpha\rangle\right],\\
|A\rangle & = & N_{-}\left[|\alpha\rangle-|-\alpha\rangle\right],
\end{array}\label{eq:one}
\end{equation}
where $N_{\pm}=\left(2\pm2\exp\left(-2|\alpha|^{2}\right)\right)^{-\frac{1}{2}}$ is
the normalization constant, and $|\alpha\rangle$ is the usual single
mode coherent state which can be expressed in the Fock state basis
as $|\alpha\rangle=\exp\left(-\frac{|\alpha|^{2}}{2}\right)\sum_{n=0}^{\infty}\frac{\alpha^{n}}{\sqrt{n!}}|n\rangle$,
where $|n\rangle$ is the Fock state. Further, for a large value of
the average photon number $|\alpha|^{2},$ we obtain $N_{\pm}=\frac{1}{\sqrt{2}},$
and thus $|S\rangle$ and $|A\rangle$ reduce to equal superposition
states of two opposite coherent states. Often, symmetric and asymmetric
cat states are referred to as the even and odd coherent states, respectively
for the obvious reason that $|S\rangle$ ($|A\rangle$) contains only
even (odd) number of photons \cite{Agarwal2013quantum}.}{\large \par}

Symmetric and asymmetric cat states of the form similar to
what is described through Eq. (\ref{eq:one}) have been realized in
various physical systems (\cite{ourjoumtsev2007generation,cochrane1999macroscopically,poyatos1996quantum} and references therein). {However,
in the present work, we would restrict our discussion to the optical
implementations only. Now, we wish to create a superposition state of symmetric
and asymmetric cat states. To do so, we may shift (displace) the symmetric
cat state in the phase space by a small amount $\beta$ in a direction
orthogonal to the orientation of the symmetric cat state in phase
space. Physically, this shifting can be done by driving the symmetric
cat state for time $t$ by a classical driving force (having a complex
amplitude $\beta$) described by the following Hamiltonian (in the
interaction picture) \cite{cochrane1999macroscopically}
\[
H_{D}=\hbar\left(\beta a^{\dagger}+\beta^{*}a\right).
\]
For the convenience, if we consider $\alpha$ and $\beta$ to be real,
then under influence of this driving force, symmetric cat state would
evolve to a quantum state 
\begin{equation}
\begin{array}{lcl}
|\psi\rangle & = & \exp\left(-i\frac{H_{D}t}{\hbar}\right)|S\rangle\\
 & \approx & \cos(\alpha\beta t)|S\rangle-i\sin(\alpha\beta t)|A\rangle\\
 & = & \exp(-\alpha\beta t)\left[|\alpha\rangle+\exp(2\alpha\beta t)|-\alpha\rangle\right]\\
 & \approx & |\alpha\rangle+\exp(i\phi)|-\alpha\rangle
\end{array}\label{eq:fprmation-of-state}
\end{equation}
}where $\phi=2\alpha\beta t$ \cite{cochrane1999macroscopically}.
This state can be viewed either as a superposition of
two opposite coherent states or equivalently as a shifted symmetric
cat state that reduces to symmetric cat state, Yurke-Stoler coherent
state, asymmetric cat state for $\phi=0,\,\phi=\frac{\pi}{2}$
and $\phi=\pi$, respectively. Further, it can be expressed in the normalized
form as 
\begin{equation}
|\alpha,\phi\rangle=\mathcal{N}_{m}^{\text{-1/2}}\left(|\alpha\rangle+e^{i\phi}|\text{\textminus}\alpha\rangle\right),\label{eq:super-position-state}
\end{equation}
where $\mathcal{N}_{m}=2+\exp\left(\text{\textminus}2|\alpha|^{2}\right)\left(e^{i\phi}+e^{\text{\textminus}i\phi}\right)$.

This is an interesting quantum state that can be produced via the
above mentioned process involving classical field. It is also interesting
because, if we can obtain an analytic expression for the witnesses
of higher-order nonclassicality for this state, then in the limiting
cases the same expressions will be of use for the study of higher-order
nonclassical properties of the symmetric cat state or even coherent
state ($\phi=0$), asymmetric cat state or odd coherent state ($\phi=\pi$)
and Yurke-Stoler coherent state ($\phi=\frac{\pi}{2}$). 

Here it may be apt to note that the witnesses of nonlcassicality are
often expressed in terms of moments of annihilation and creation operators
\cite{miranowicz2010testing,shchukin2005nonclassical}. To obtain
the witnesses of higher-order nonclassicality, we may use the following
analytic expression which has been obtained by Ren et al.,\cite{ren2017non}
in a recent work related to lower-order nonclassicalities of $|\alpha,\phi\rangle$

\begin{equation}
\mathcal{H}_{1}\left(k,l\right)\text{\ensuremath{\equiv}}\langle\alpha,\phi|a^{\dagger k}a^{l}|\alpha,\phi\rangle=[1+(\text{\textminus}1)^{k+l}]\alpha^{\text{\ensuremath{\star}}k}\alpha^{l}+\exp(\text{\textminus}2|\alpha|^{2})[(\text{\textminus}\alpha^{\text{\ensuremath{\star}}})^{k}\alpha^{l}e^{\text{\textminus}i\phi}+\alpha^{\star k}(\text{\textminus}\alpha)^{l}e^{i\phi}].\label{eq:hkl}
\end{equation}
In what follows, we would express the witnesses of different features
of higher-order nonclassiality using $\mathcal{H}_{1}\left(k,l\right)$
and report their compact analytic expressions in most cases. We will
also illustrate the variation of the witnesses of higher-order nonclassicality
with various parameters with a specific focus on the limiting cases
of shifted symmetric and cat state. Before, we do that it will
be apt to go beyond the discrete superposition of coherent states
and describe a continuous superposition of coherent states in the
following subsection.

\subsection{One-dimensional continuous superposition of coherent states}

One-dimensional continuous superposition of coherent states have been
studied since long. Specifically, continuous superposition of the
following form has been studied in Refs. \cite{janszky1990squeezing,buvzek1991origin,adam1990gaussian,adam1991amplitude,foldesi1993antisymmetric}
and nicely reviewed in Refs. \cite{buvzek1995quantum,buvzek1992superpositions}

\begin{equation}
|\xi\rangle=C_{F}\stackrel[-\infty]{\infty}{\int}F\left(\alpha,\xi\right)|\alpha\rangle d\alpha,\label{eq:CVstate1}
\end{equation}
where the complex parameter $\xi=\left|\xi\right|e^{i\theta}$ and
$C_{F}$ is the normalization factor and is defined as 

\begin{equation}
C_{F}^{-2}=\stackrel[-\infty]{\infty}{\iint}F\left(\alpha,\xi\right)F\left(\alpha',\xi\right)\exp\left[-\left(\alpha-\alpha'\right)/2\right]d\alpha d\alpha'=\frac{2\pi\xi}{\sqrt{1-\xi^{2}}}.\label{eq:cvstate2}
\end{equation}
In the earlier works \cite{janszky1990squeezing,buvzek1991origin,adam1990gaussian,adam1991amplitude,foldesi1993antisymmetric},
it had been shown that with a properly selected weight function $F\left(\alpha,\xi\right)$,
one can considerably enhance the degree of squeezing \cite{buvzek1995quantum}.
For the following choice of weight function $F\left(\alpha,\xi\right)$
\begin{equation}
F\left(\alpha,\xi\right)=\exp\left(-\frac{1-\xi}{2\xi}\alpha^{2}\right),
\end{equation}
(i.e., when $F\left(\alpha,\xi\right)$ is chosen to be Gaussian),
quantum state described in Eq. (\ref{eq:CVstate1}) is known to reduce
to the squeezed vacuum state $|\xi\rangle=C_{F}\stackrel[-\infty]{\infty}{\int}F\left(\alpha,\xi\right)\hat{D}(\alpha)|0\rangle d\alpha=\hat{S}(\xi)|0\rangle,$
where $\hat{D}(\alpha)$ and $\hat{S}(\xi)$ are the displacement
and squeezing operators, respectively \cite{buvzek1995quantum}. In
what follows, we would study the higher-order nonclassical properties
of this particular state (i.e., squeezed vacuum state) as a representative
state of the set of possible quantum sates that can be viewed as one-dimensional
continuous superposition of coherent states. Lower-order nonclassical
properties of this state has been studied earlier in Ref. \cite{buvzek1991origin}.
In order to study the higher-order nonclassical properties of this
state using moment-based criteria of nonclssicality, it would be helpful
to obtain an expression for $\langle\xi|a^{\dagger k}a^{l}|\xi\rangle$ in a manner similar
to Eq. (\ref{eq:hkl}) which was obtained for the discrete superposition
of coherent states. Such an expression for the continuous superposition
of coherent states of our interest can be written as follows

\begin{equation}
\langle\xi|a^{\dagger k}a^{l}|\xi\rangle=\mathcal{H}_{2}(k,l)=C_{F}\stackrel[-\infty]{\infty}{\iint}F\left(\alpha,\xi\right)F\left(\alpha',\xi\right)\exp\left[-\left(\alpha-\alpha'\right)/2\right]\alpha^{k}\left(\alpha'\right)^{l}d\alpha d\alpha'.\label{eq:Hkl2}
\end{equation}

In what follows, we would observe that Eq. (\ref {eq:Hkl2}) Would be useful in investigating various higher-order nonclassical properties of $|\xi\rangle$
\section{Witnesses of higher-order nonclassicality and their variation \label{sec:Witnesses-of-higher}}

Witnesses of nonclassicality can be expressed as moments of annihilation
and creation operators. In the next section, we will use moment-based
criteria to find the signatures of higher-order nonclassical features present
in the discrete and continuous superposition of coherent states described
above. To do so, we will use the above expressions of $\mathcal{H}_{j}(k,l)$
where $j\in\{1,\,2\}$. For the convenience of the study, various
criteria of higher-order nonclassicality are stated below in their
standard form as well as in terms of $\mathcal{H}_{j}(k,l)$.

\subsection{Higher-order antibunching}

Phenomenon of the antibunching is closely related to photon statistics. In fact, it can be lucidly described as a phenomenon that ensures that in an incident beam of radiation, the probability of getting
two photons simultaneously is less than the probability of
getting them separately (one-by-one). The natural tendency of an ordinary light
source is to emit photons in a bunched state, but there are some nonclassical sources of light which emit photons
in the antibunched state. Being nonclassical, they are of extreme interest. Experimentally, antibunching  was first demonstrated by Hanbury-Brown
and Twiss in their pioneering experiment\cite{brown1956correlation}.
Generalizing the idea of lower-order antibunching, the notion of HOA
was introduced by Lee in 1990 \cite{lee1990many}, when he proposed the first criterion of  HOA using the theory of majorization. Later, Lee's
criterion was modified by Ba An \cite{an2002multimode} and Pathak and Garcia
\cite{pathak2006control}, respectively. In this article we would 
follow Pathak and Garcia's criterion to observe $l$th order antibunching. Specifically, in what follows following criterion for HOA 
(which is now known as Pathak and Garcia criterion) is used 

\begin{equation}
D(l)=\left\langle N^{(l+1)}\right\rangle -\left\langle N\right\rangle ^{l+1}=\langle a^{\dagger l+1}a^{l+1}\rangle-\langle a^{\dagger}a\rangle^{l+1}<0.\label{eq:ho21}
\end{equation}
where $N=a^{\dagger}a$ and $N^{(l+1)}=a^{\dagger l+1}a^{l+1}$ are
the number operator and factorial moment, respectively. It's easy to observe that we can express the criterion described by  Eq. (\ref{eq:ho21}) in a convenient form as follows

\begin{equation}
D(l)=\mathcal{H}_{j}\left(l+1,\,l+1\right)-\left(\mathcal{H}_{j}\left(1,\,1\right)\right)^{l+1}<0.\label{eq:HOA-simple}
\end{equation}
For the shifted symmetric cat state, we have $j=1$, and in that case, we can use  Eqs (\ref{eq:hkl}) and (\ref{eq:HOA-simple}) to obtain the witness of HOA (i.e., $D(l)$).  The obtained expression of $D(l)$ is illustrated in Fig. \ref{HOA and HOSPS} a-c, where the negative parts of the curves depict the presence of HOA of the $l$th order. It can be easily observed that the HOA is observed only
for a certain range of $\phi$ (see Fig. \ref{HOA and HOSPS} c ). Further, it is observed that the depth of nonlciassicality witness is more for the higher values of $l$, which is consistent with the earlier observations \cite{verma2008higher}. Interestingly, in the cases where HOA is found, it is observed that  with the increase in the average photon number the depth of nonclassicality increases in the beginning, but reduces after a point (cf. Fig. \ref{HOA and HOSPS} a-b). Further, it is clear from Fig. \ref{HOA and HOSPS} c that HOA is not
observed for even coherent state ($\phi=0$)
and Yurke-Stoler coherent state ($\phi=\frac{\pi}{2}$), but it
is observed for odd coherent state ($\phi=\pi$). Thus, in brief, we have seen that HOA can be observed in the discrete superposition of coherent states studied here. In contrast, while we performed the similar exercise using Eqs (\ref{eq:Hkl2}) and (\ref{eq:HOA-simple}), we failed to locate any signature of HOA for the quantum state formed by one-dimensional continuous superposition of coherent states. This motivated us to look into the possibility of observing higher-order nonclassicality in the one-dimensional continuous superposition of coherent states by using the other witnesses of higher-order nonclassicality.

\subsection{Higher-order sub-Poissonian photon statistics }

\begin{figure}
\centering

\subfigure[]{\includegraphics[scale=0.55]{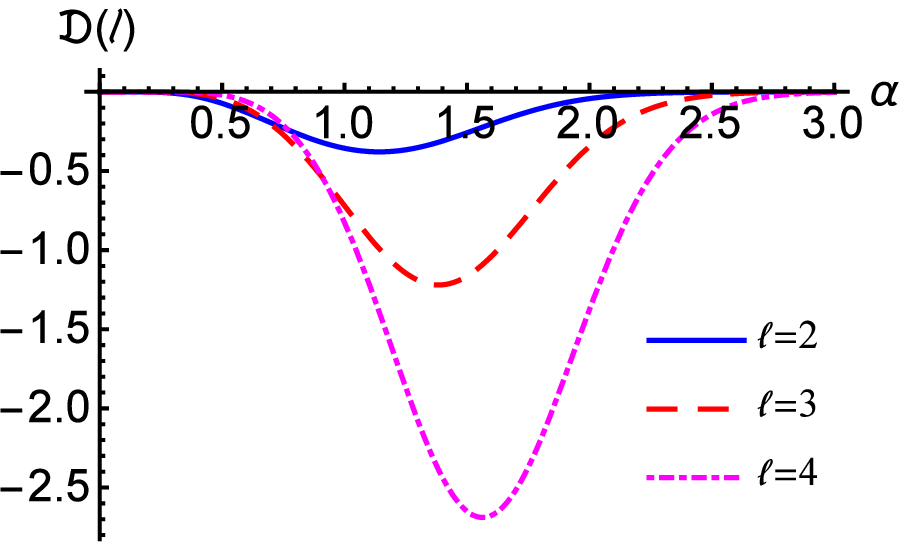}} \quad \subfigure[]{\includegraphics[scale=0.55]{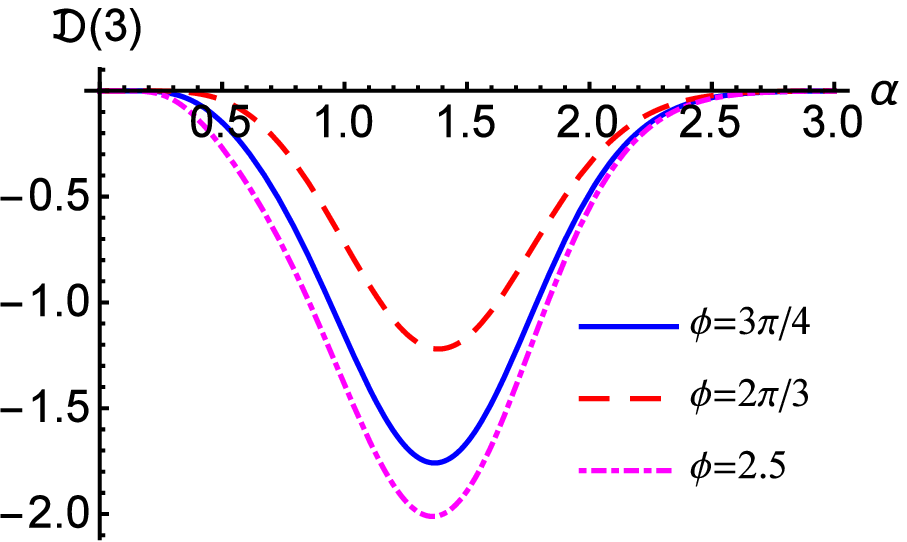}}\quad \subfigure[]{\includegraphics[scale=0.55]{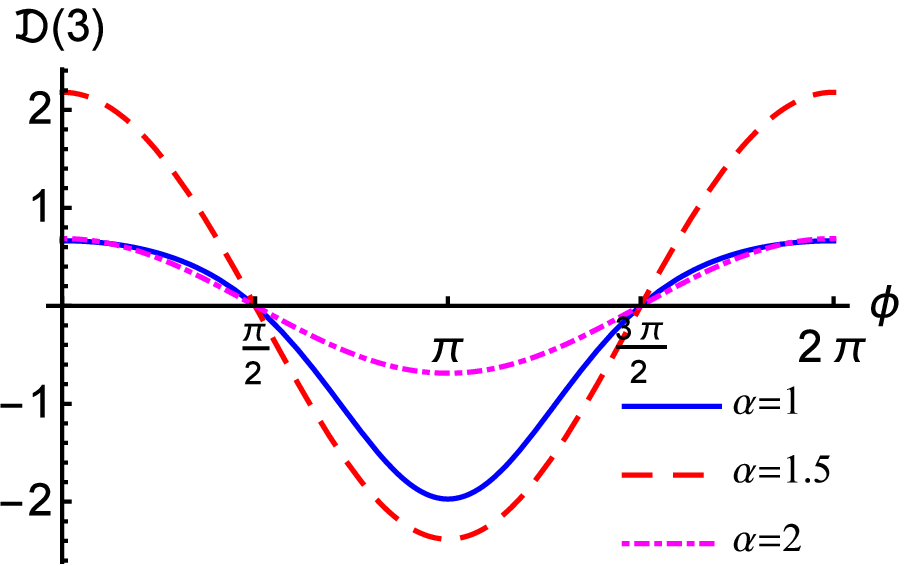}}\\
\subfigure[]{\includegraphics[scale=0.55]{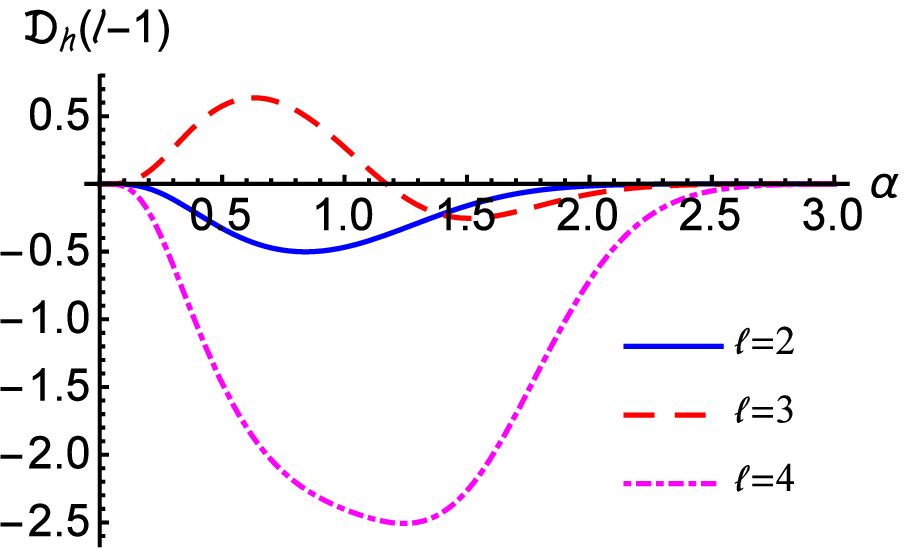}}\quad
\subfigure[]{\includegraphics[scale=0.55]{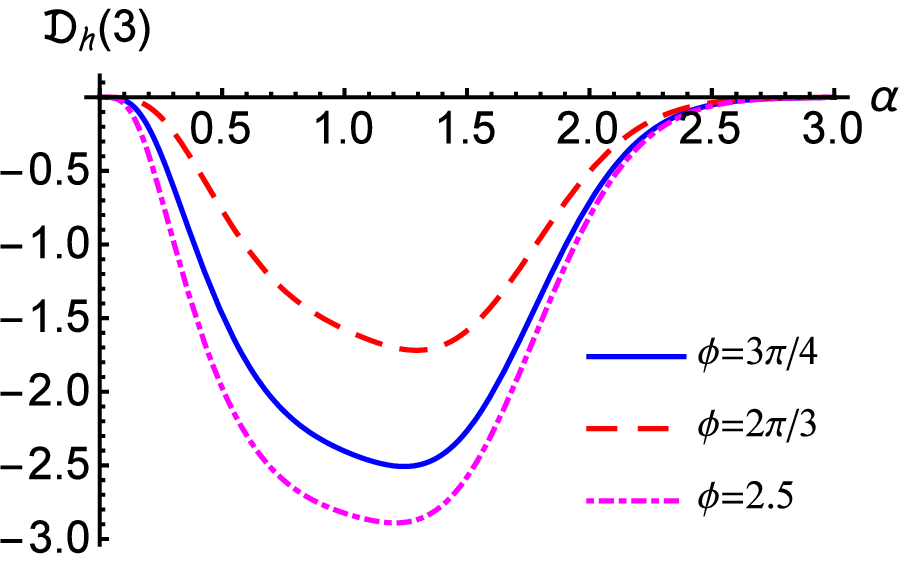}}\quad\subfigure[]{\includegraphics[scale=0.55]{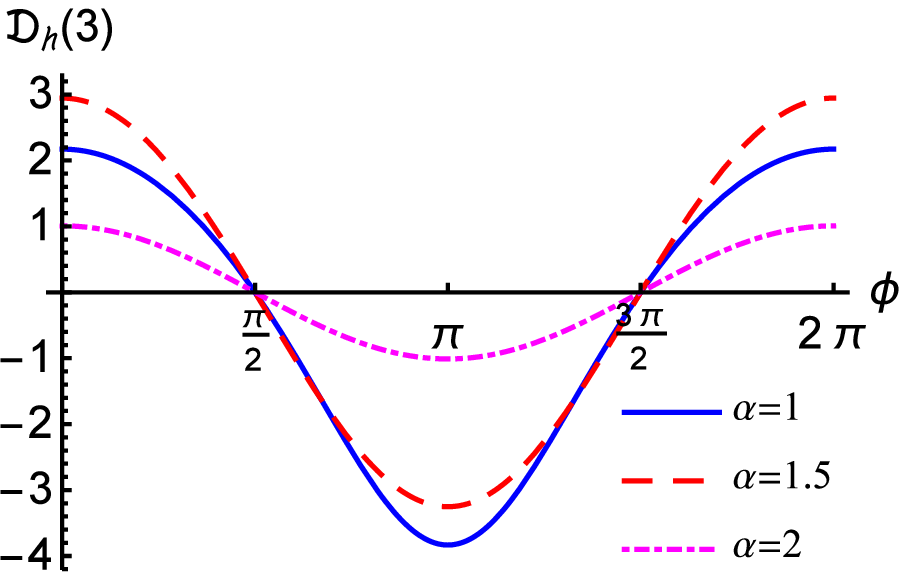}}\\
\caption{\label{HOA and HOSPS}(Color online) (a)-(c) HOA, (d)-(f) HOSPS for the discrete
superposition of coherent states. (a) and (d) show increase in the depth of witness of nonclassicality with order and coherent state parameter. Dependence of the observed nonclassicality on the  state parameters are shown in (b), (c), (e) and (f).}
\end{figure}

Higher-order nonclassical feature associated with the photon statistics of a quantum state of radiation field is usually studied through the witness of HOSPS \cite{prakash2006higher,verma2010generalized}. Here it may be noted that HOSPS is the higher-order analogue of the frequently investigated sub-Poissonian photon statistics \cite{agarwal1974quantum}.  The generalized criterion
to observe HOSPS is given by \cite{verma2010generalized}

\begin{equation}
\begin{array}{lcccc}
D_{h}(l-1) & = & \sum\limits _{r=0}^{l}\sum\limits _{k=0}^{r}S_{2}(r,\,k)\,^{l}C_{r}\,\left(-1\right)^{r}D(k-1)\langle N\rangle^{l-r} & < & 0,\end{array}\label{eq:hosps22}
\end{equation}
where $S_{2}(r,\,k)$ is the Starling number of second kind. For our convenience, the above
inequality can written in terms of $\mathcal{H}_{j}(k,l)$ in the 
the following form

\begin{equation}
\begin{array}{lcccc}
D_{h}(l-1) & = & \sum\limits _{r=0}^{l}\sum\limits _{k=0}^{r}S_{2}(r,\,k)\,^{l}C_{r}\,\left(-1\right)^{r}\left\{ \mathcal{H}_{j}\left(k,\,k\right)-\left(\mathcal{H}_{j}\left(1,\,1\right)\right)^{k}\right\} \left(\mathcal{H}_{j}\left(1,1\right)\right)^{l-r} & < & 0.\end{array}\label{eq:hosps_final}
\end{equation}
The existence HOSPS is investigated for both discrete and continuous superposition
of coherent states using  Eqs. (\ref{eq:hkl}), (\ref{eq:Hkl2}) and (\ref{eq:hosps_final}). The obtained results are explicitly illustrated in Fig.
  \ref{HOA and HOSPS} d-f for discrete superposition of coherent states and in Fig. \ref{HOA-contitnuous} a for the continuous superposition of coherent states. The negative parts of the plots depict the situation where quantum interference in the phase space leads to higher-order nonclassical photon statistics (in particular HOSPS). In Fig. \ref{HOA and HOSPS} f, it is easily observed that there exists
a nonclassical region i.e., a range of $\phi$ in which the curves are negative. This is logically expected and observed in the context of HOA, too. In addition, Fig.
\ref{HOA and HOSPS} f also illustrates that the HOSPS is not observed for  even coherent state ($\phi=0$) and Yurke-Stoler coherent
state ($\phi=\frac{\pi}{2}$), but it is observed for  odd coherent state ($\phi=\pi$). Similarly, the quantum inference caused by continuous superposition of coherent states also leads to HOSPS as shown 
in Fig. \ref{HOA-contitnuous} a. Further, comparing Fig. \ref{HOA and HOSPS} a and d  for $l=3$ we can see that HOSPS is absent for small values of average photon number (cf. Fig. \ref{HOA and HOSPS} d), but HOA is found to be present for the same values of average photon number (cf. Fig. \ref{HOA and HOSPS} a). In contrast, HOA is not observed for the one-dimensional continuous superposition of coherent states, but HOSPS is observed for the same state (cf. Fig. \ref{HOA-contitnuous} a). These observations establish that HOA and HOSPS are independent phenomena and the existence of one does not ensure the existence of the other.  This is in consistency with our earlier works (\cite{alam2017higher,thapliyal2017comparison} and references therein). 

\subsection{Higher-order squeezing}

Non-commutativity of the field quadrature operators (which are usually expressed as functions of annihilation and creation operators) lead to Heisenberg uncertainty relations. Usually, for the coherent states (in the infinite dimensional scenario), the product of the fluctuations in two field quadrature becomes minimum and the fluctuations in each quadrature become equal. For the lower-order  squeezing, variance in one of the field quadrature (defined as a linear combination of annihilation and creation operator) reduces below the coherent state value at the cost of enhanced fluctuations in the other quadrature. If we now look at the quantum fluctuations of the field quadrature using higher-order moments (higher than $n=2$ which corresponds to variance) and claim that $n^{\rm{th}}$ order squeezing will be observed when $n^{\rm{th}}$ moment of a field quadrature operator will be less than the corresponding coherent state value, then we would obtain the condition of Hong-Mandel type HOS as follows \cite{hong1985higher,verma2010generalized} 

\begin{equation}
S_{HM}(n)=\frac{\langle(\Delta X)^{n}\rangle-\left(\frac{1}{2}\right)_{\frac{n}{2}}}{\left(\frac{1}{2}\right)_{\frac{n}{2}}}<0,\label{eq:Hong-def}
\end{equation}
where $(x)_{n}$ is conventional Pochhammer symbol. The above inequality can be rewrite as
\begin{equation}
\begin{array}{lcl}
\langle(\Delta X)^{n}\rangle & < & \left(\frac{1}{2}\right)_{\frac{n}{2}}=\frac{1}{2^{\frac{n}{2}}}(n-1)!!\end{array},\label{eq:Hong-def2} 
\end{equation}
where

\begin{equation}
\begin{array}{lcl}
\langle(\Delta X)^{n}\rangle & = & \sum\limits _{r=0}^{n}\sum\limits _{i=0}^{\frac{r}{2}}\sum\limits _{k=0}^{r-2i}(-1)^{r}\frac{1}{2^{\frac{n}{2}}}(2i-1)!!\,^{r-2i}C_{k}\,^{n}C_{r}\,^{r}C_{2i}\langle a^{\dagger}+a\rangle^{n-r}\langle a^{\dagger k}a^{r-2i-k}\rangle.\end{array}\label{eq: Hong-mandel}
\end{equation}
The expression for the $n$th order moment i.e., Eq. (\ref{eq: Hong-mandel})  can now be rewritten in a convenient form   in
terms of $\mathcal{H}_{j}(k,l)$ as 

\begin{equation}
\begin{array}{lcl}
\langle(\Delta X)^{n}\rangle & = & \sum\limits _{r=0}^{n}\sum\limits _{i=0}^{\frac{r}{2}}\sum\limits _{k=0}^{r-2i}(-1)^{r}\frac{1}{2^{\frac{n}{2}}}(2i-1)!!\,^{r-2i}C_{k}\,^{n}C_{r}\,^{r}C_{2i}\left\{ \mathcal{H}_{j}\left(1,0\right)+\mathcal{H}_{j}\left(0,1\right)\right\} ^{n-r}\\
 & \times & \mathcal{H}_{j}\left(k,\,r-2i-k\right).
\end{array}\label{eq:Hong-Mandel2}
\end{equation}

Another possible way to study the HOS was introduced by Hillery who introduced HOS as a phenomenon where variance of amplitude-powered field quadrature reduces below the coherent state value.  According to the Hillery, the $l$th order amplitude
powered field quadrature are defined as \cite{hillery1987amplitude}

\begin{equation}
\begin{array}{lcl}
Y_{1,a} & = & \frac{a^{l}+a^{\dagger l}}{2},\\
Y_{2,a} & = & -i\frac{a^{l}-a^{\dagger l}}{2},
\end{array}\label{eq:19}
\end{equation}
where the quadratures, $Y_{1,a}$ and $Y_{2,a}$, are non-commuting
in nature. The HOS criteria involving these non-commuting field  quadrature
can be obtained as 
\begin{equation}
\mathcal{A}_{i,a}=\left(\Delta Y_{i,a}\right)^{2}-\frac{1}{2}\left|\left\langle \left[Y_{1,a},Y_{2,a}\right]\right\rangle \right|<0,\label{eq:n-th-squeezing}
\end{equation}
where $i\in\{1,2\}.$ In this paper, we have used the Hillery's criterion
for amplitude square squeezing, i.e., in Eq. (\ref{eq:n-th-squeezing}),
we choose $l=2$ and have reduced the criterion of HOS given in Eq. (\ref{eq:20})
to

\begin{equation}
\begin{array}{lcl}
\mathcal{A}_{i,a} & = & \left(\Delta Y_{i,a}\right)^{2}-\left|\langle N_{a}+\frac{1}{2}\rangle\right|<0,\end{array}\label{eq:20}
\end{equation}
where $N_{a}$ corresponds to the number operator of the mode $a$.
A few steps of computation have yielded the condition of Hillery type HOS for
$l=2$ as follows 
\begin{equation}
\begin{array}{lcl}
\left[\begin{array}{c}
\mathcal{A}_{1,a}\\
\mathcal{A}_{2,a}
\end{array}\right] & = & \frac{1}{4}\left[\pm\langle\left(a^{\dagger}\right)^{4}\rangle\pm\langle\left(a\right)^{4}\rangle+2\langle\left(a^{\dagger2}a^{2}\right)\rangle\mp\left(\langle\left(a^{\dagger}\right)^{2}\rangle\pm\langle\left(a\right)^{2}\rangle\right)^{2}\right]<0,\end{array}\label{eq:21}
\end{equation}
where the upper and lower sign in Eq. (\ref{eq:21}) correspond to $\mathcal{A}_{1,a}$
and $\mathcal{A}_{2,a}$, respectively. As before, we can rewrite Eq. (\ref{eq:21})  in a more convenient form in terms of $\mathcal{H}_{j}(k,l)$ as 

\begin{equation}
\begin{array}{lcl}
\left[\begin{array}{c}
\mathcal{A}_{1,a}\\
\mathcal{A}_{2,a}
\end{array}\right] & = & \frac{1}{4}\left[\pm\mathcal{H}_{j}\left(4,0\right)\pm\mathcal{H}_{j}\left(0,4\right)+2\mathcal{H}_{j}\left(2,2\right)\mp\left(\mathcal{H}_{j}\left(2,0\right)\pm\mathcal{H}_{j}\left(0,2\right)\right)^{2}\right]<0\end{array}.\label{eq:21-2}
\end{equation}
The above expression is further simplified for continuous superposition
of coherent states of our interest to yield 

\begin{equation}
\begin{array}{lcl}
\mathcal{A}_{1,a} & = & \frac{\xi^{2}\left(1+\xi^{2}\right)}{1-\xi^{2}},\\
\mathcal{A}_{2,a} & = & \frac{\xi^{2}}{\xi^{2}-1}.
\end{array}\label{eq:21-1}
\end{equation}

\begin{figure}
\centering
\subfigure[]{\includegraphics[scale=0.55]{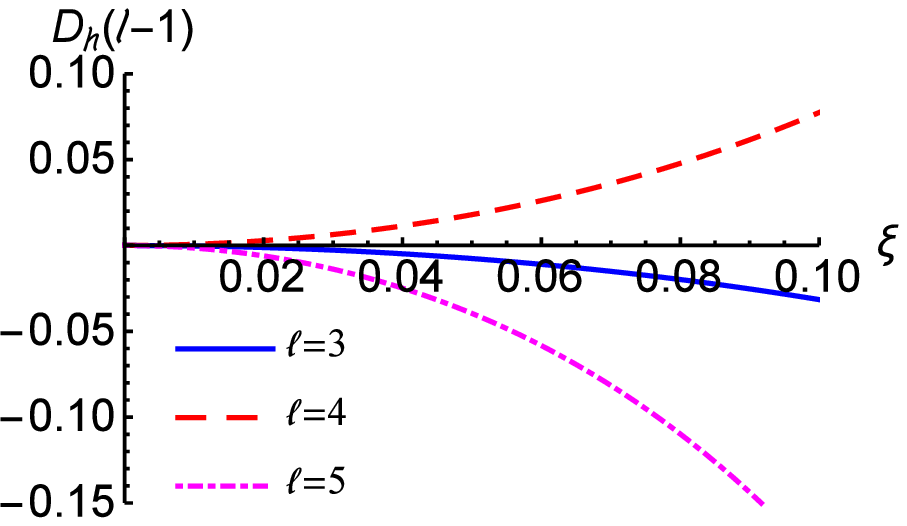}}\quad 
\subfigure[]{\includegraphics[scale=0.55]{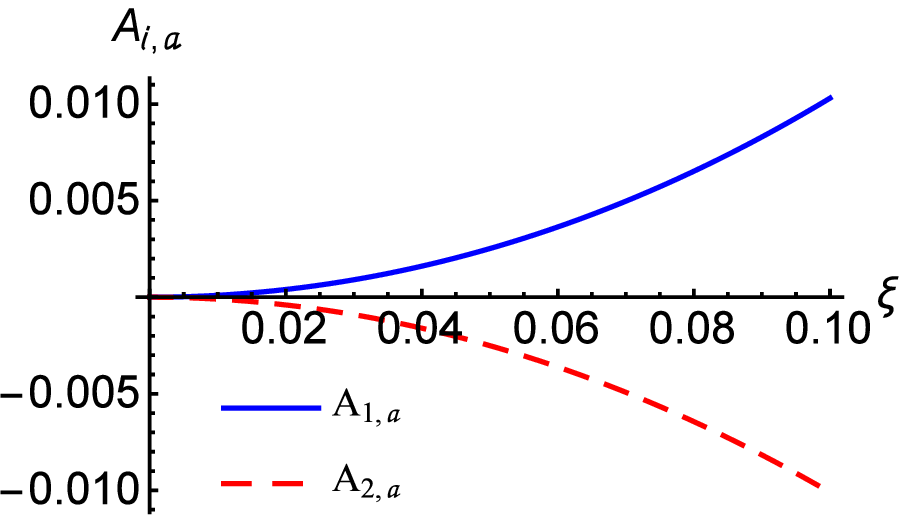}}\quad 
\subfigure[]{\includegraphics[scale=0.32]{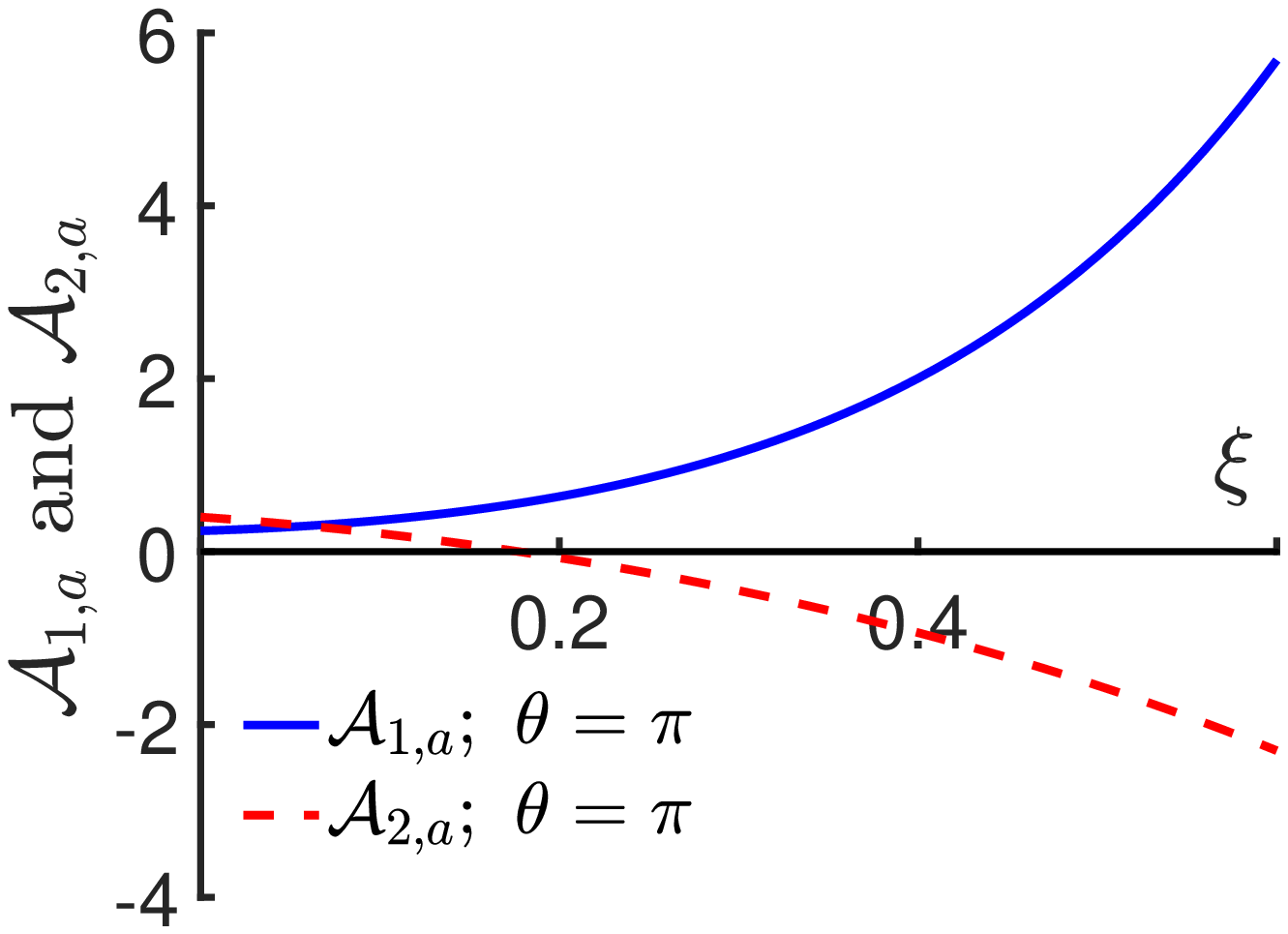}}\\
\caption{\label{HOA-contitnuous}(Color online)(a) HOSPS and (b) HOS (Hillery type) for one-dimensional
continuous superposition of coherent states with $\theta=0$. (c) HOS
(Hillery type) for a single photon added one-dimensional continuous
superposition of coherent states. }
\end{figure}

In order to investigate the possibility of observing HOS of Hillery type in the one-dimensional continuous superposition of  coherent states, right hand side of Eq. (\ref{eq:21-1}) is plotted and the
result is depicted in Fig. \ref{HOA-contitnuous} b where the presence of negative
parts establish the existence of the HOS in the one-dimensional continuous superposition
of  coherent states. In sharp contrast to it, HOS of any type is not observed in the discrete superposition
of the coherent states. However, in what follows we will show that  HOS of Hillery type can be observed for the shifted cat state after addition of photon. 

\begin{figure}
\centering
\subfigure[]{\includegraphics[scale=0.4]{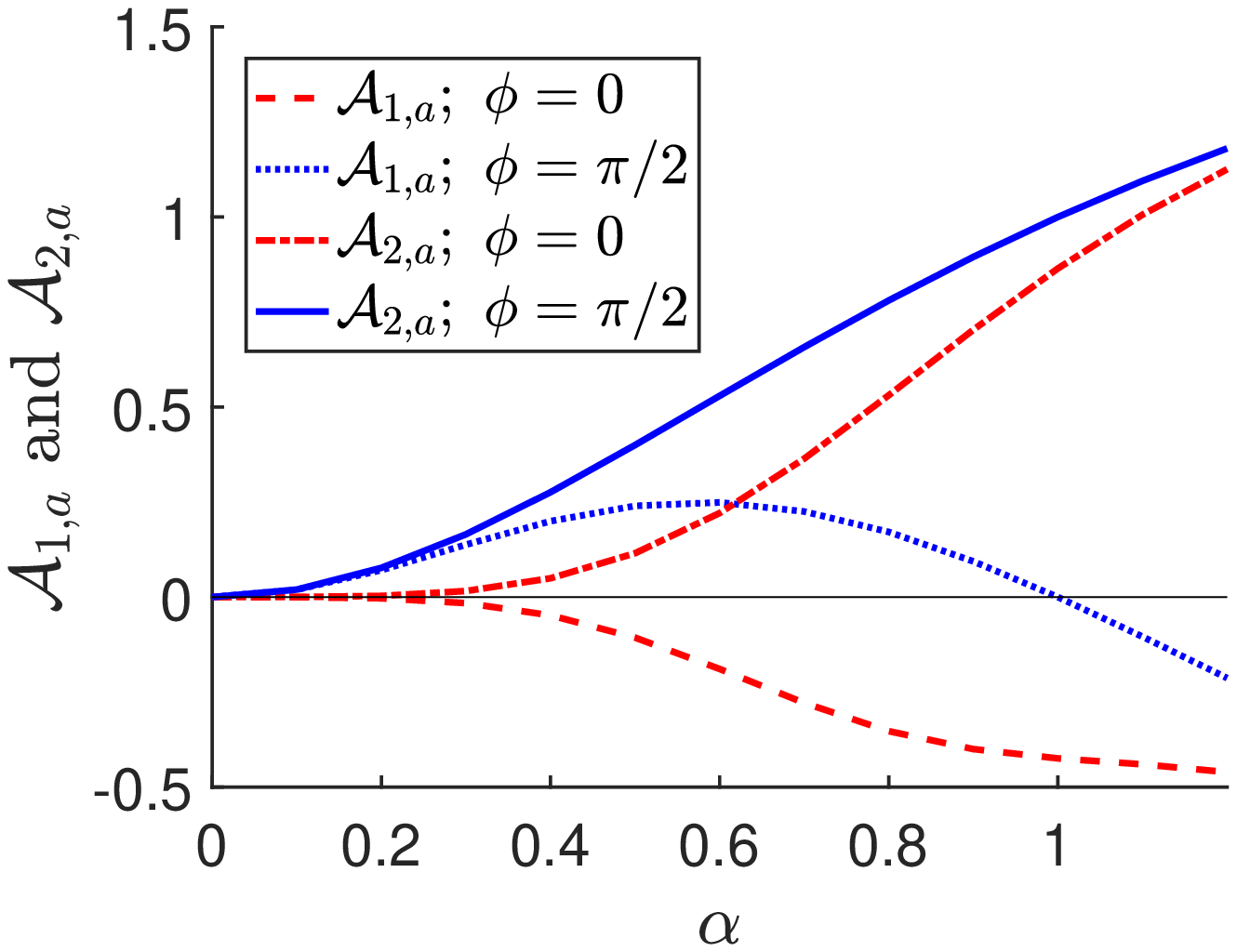}}\quad \subfigure[]{\includegraphics[scale=0.4]{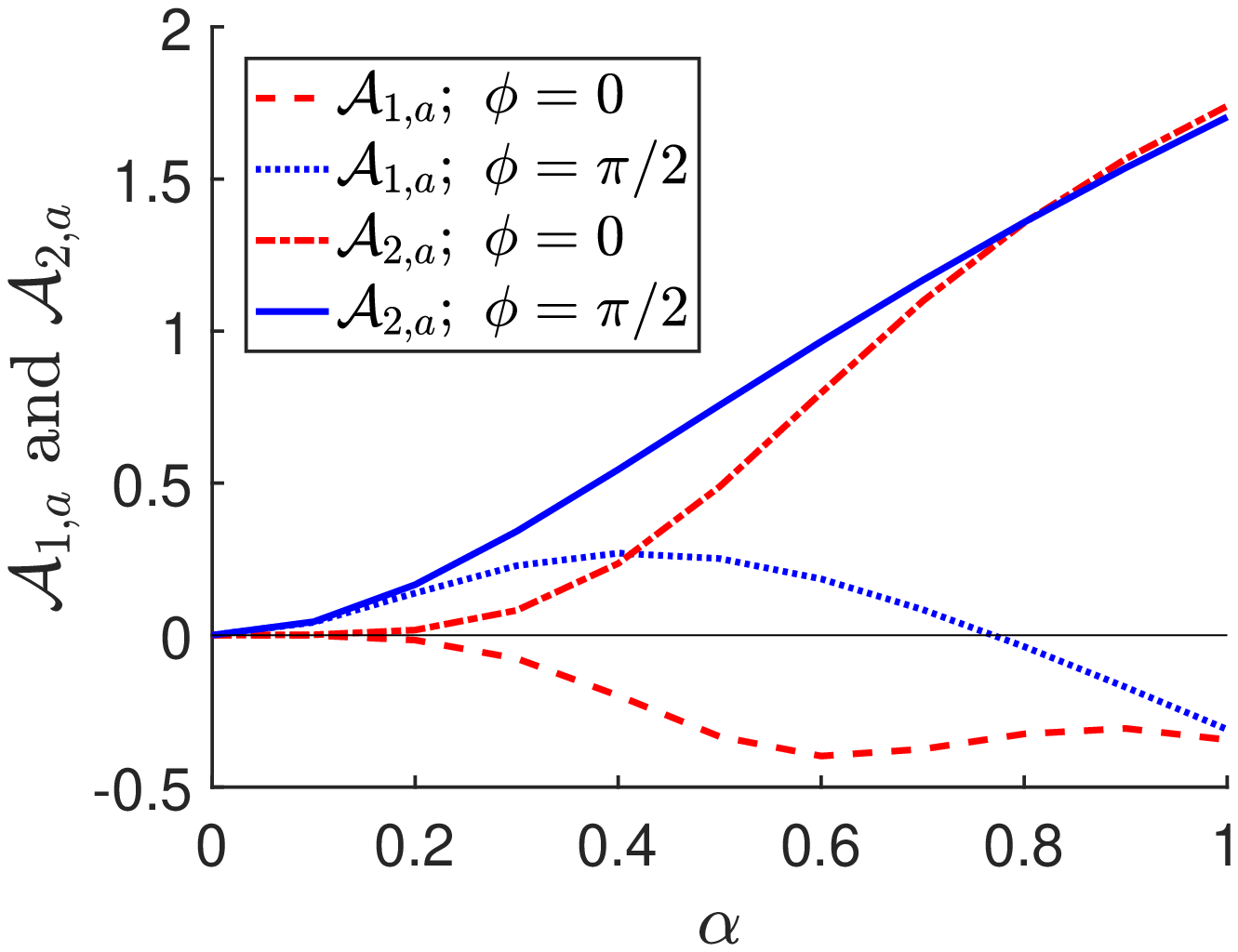}}\\
\subfigure[]{\includegraphics[scale=0.4]{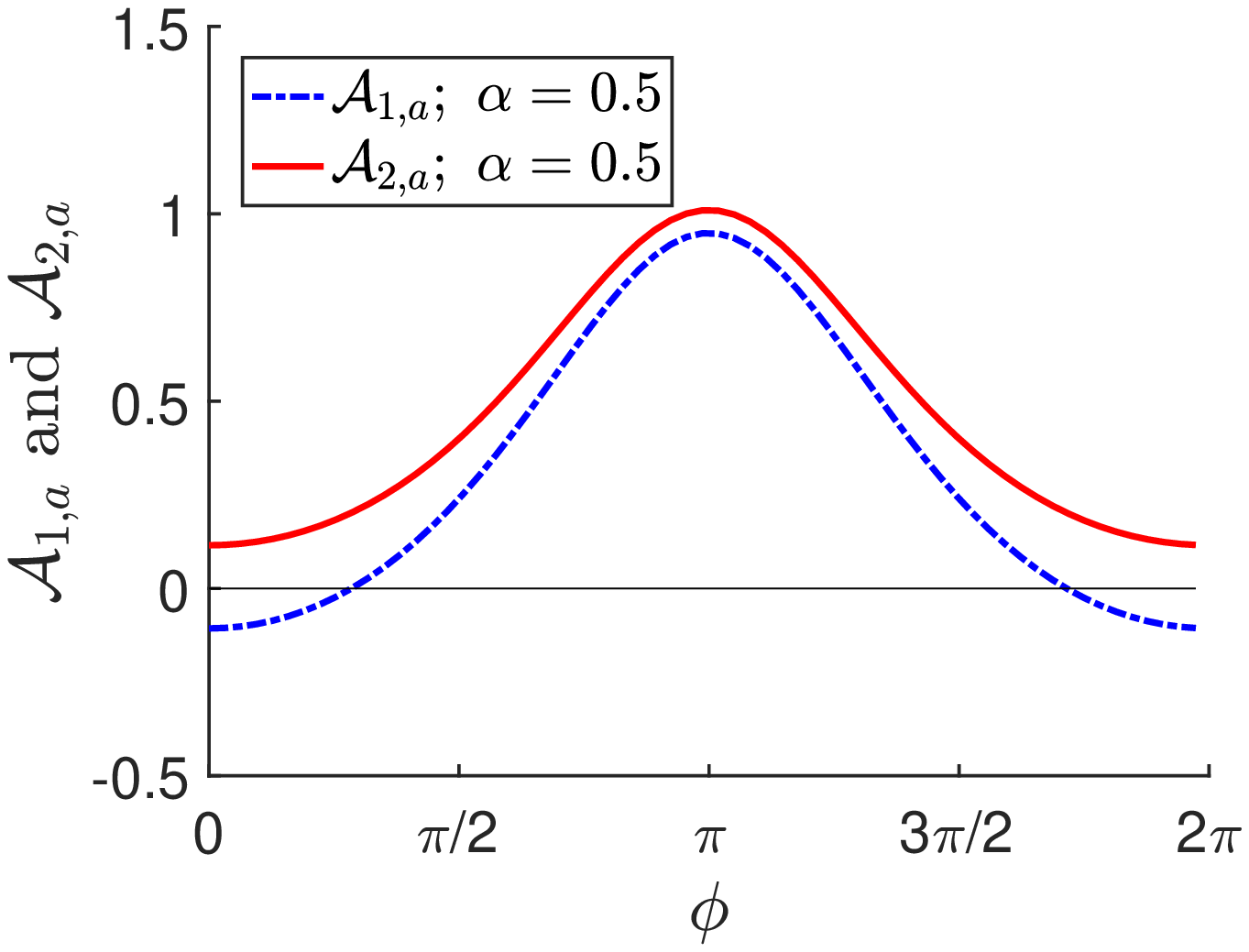}}\quad
\subfigure[]{\includegraphics[scale=0.4]{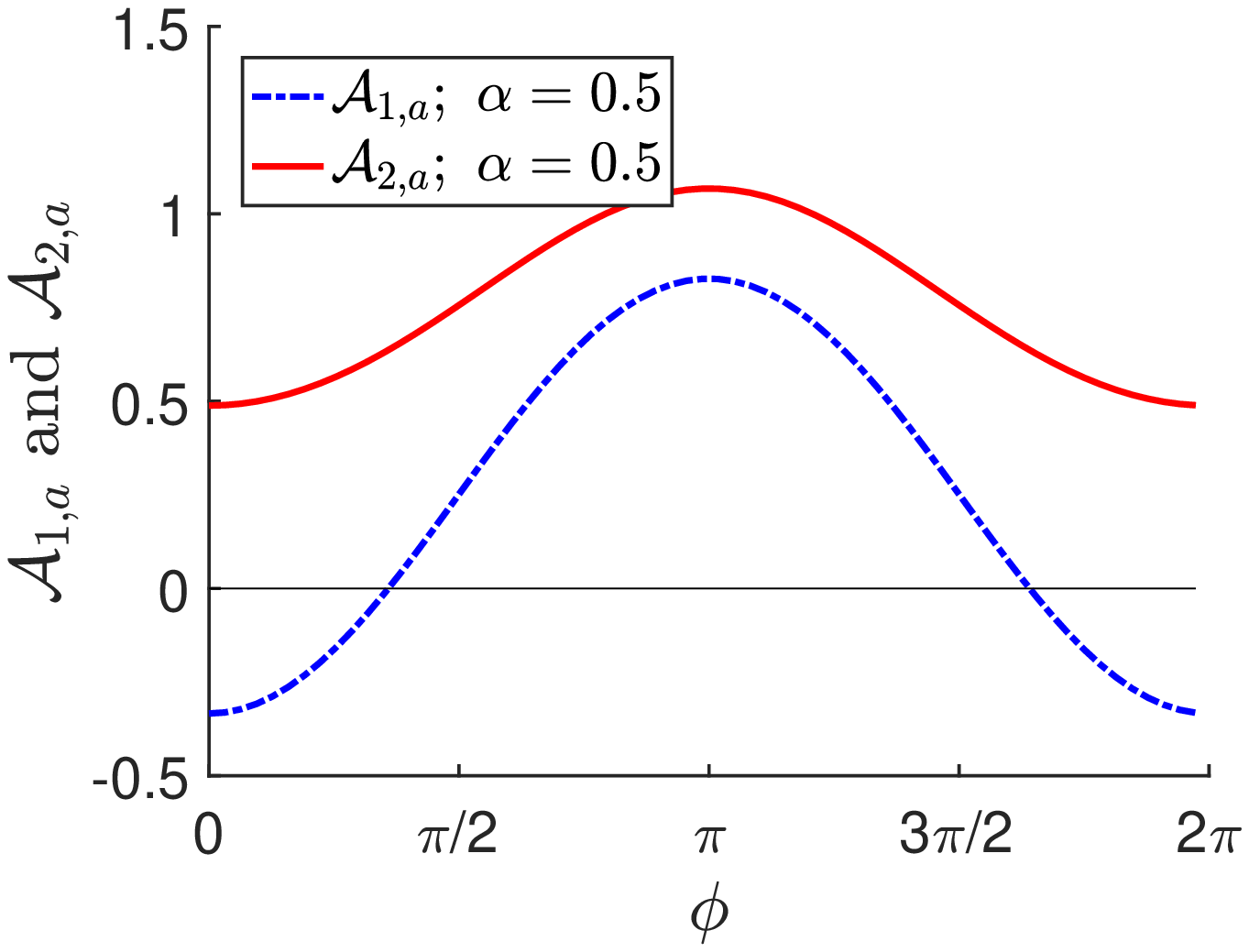}}\\
\caption{\label{fig.HOS-discrete}(Color online) Variation of $\mathcal{A}_{1,a}$ and $\mathcal{A}_{2,a}$
for single photon added discrete superposition of coherent states 
with state parameters (a) $\alpha$ and (b) $\phi$. Variation of $\mathcal{A}_{1,a}$
and $\mathcal{A}_{2,a}$ for two photon added (shown in c) and a single photon
subtracted after adding two photons (in d) discrete superposition of coherent states.}
\end{figure}

\section{Effect of photon addition and addition followed by subtraction on the higher-order nonclassical
properties\label{sec:Effect-of-photon}}

Addition of photon is known to introduce non-Gaussianity and nonclassicality in the coherent state which is classical and Gaussian. To be precise, in a pioneering work, in 1991, Agarwal and Tara introduced photon added coherent state (PACS) \cite{agarwal1991nonclassical} and shown that PACS has nonclassical properties. Later on higher-order nonclassical properties of PACS  have been studied by some of the present authors \cite{verma2008higher,verma2010higher}.  In addition to PACS, effect of photon addition, subtraction and addition followed by subtraction on various quantum states (including but not limited to squeezed coherent state \cite{thapliyal2017comparison}, thermal state\cite{zavatta2007experimental}, squeezed thermal state \cite{hu2010photon}) have been studied. The effect of the photon
addition on various quantum state has been investigated in various
literature. Specifically, nonclassicality of photon added coherent
\cite{Agarwal2013quantum,agarwal1991nonclassical}, photon added squeezed coherent
\cite{thapliyal2017comparison}, photon added thermal state \cite{zavatta2007experimental},
etc., have been investigated.

\begin{figure}
\centering
\subfigure[]{\includegraphics[scale=0.5]{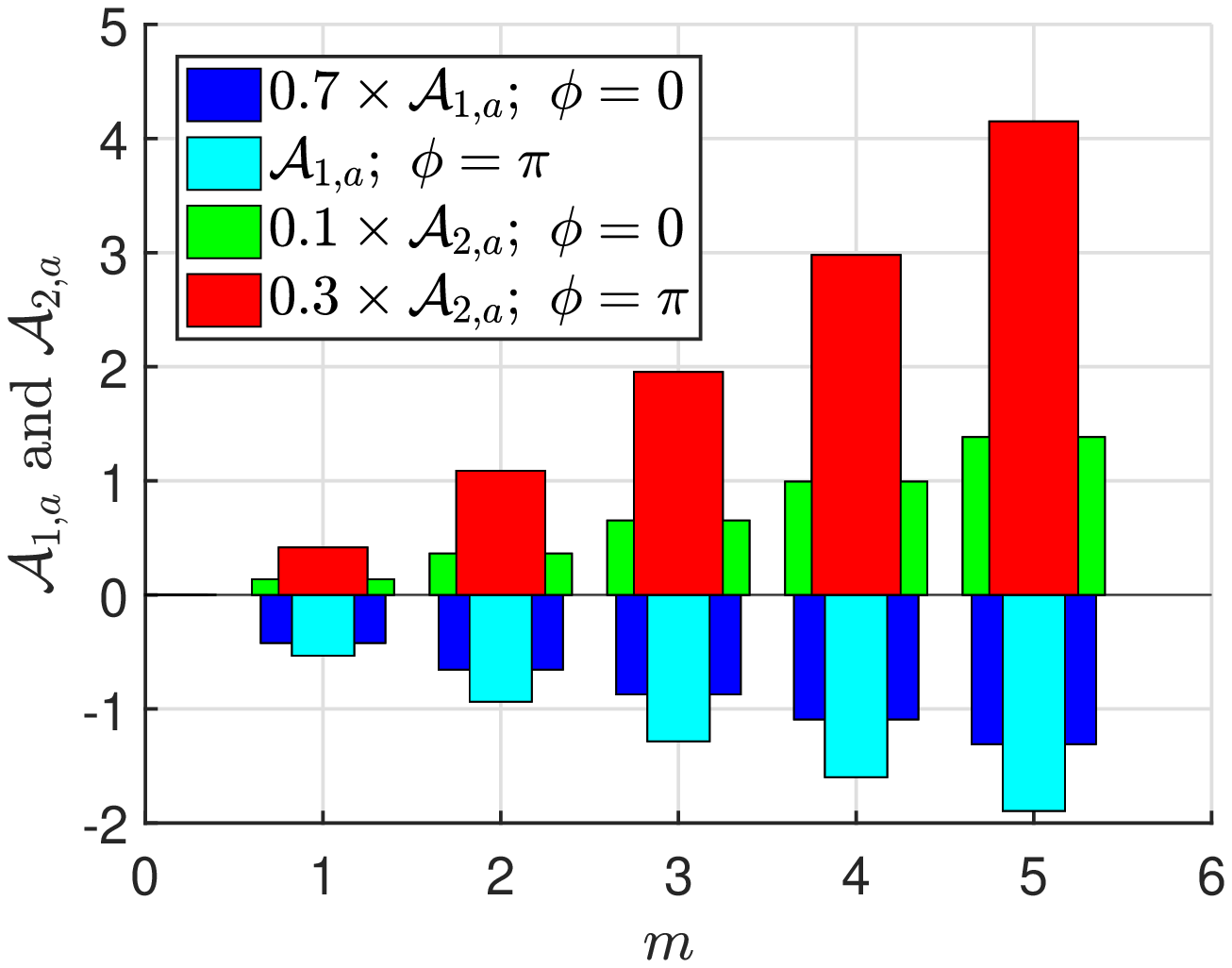}}\quad \subfigure[]{\includegraphics[scale=0.5]{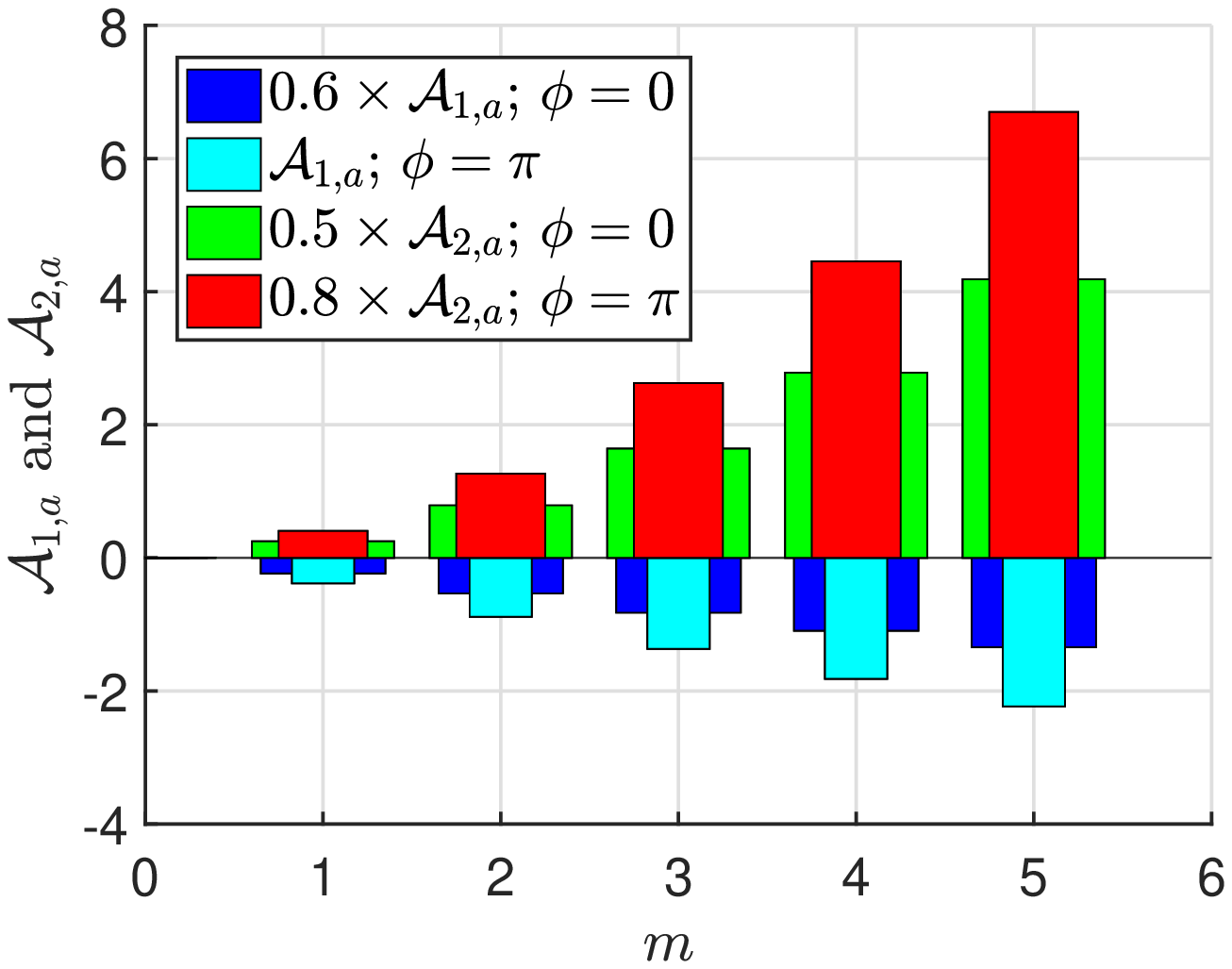}}\\
\caption{\label{fig.photon-added}(Color online) Variation of $\mathcal{A}_{1,a}$ and $\mathcal{A}_{2,a}$
with photon addition for discrete superposition of coherent states for $\alpha=1.5$ 
with (a) $m$-photon addition, (b) $m$-photon addition followed by $2$
-photon subtraction. }
\end{figure}

In the above we have mentioned that coherent state is classical, but PACS is nonclassical and thus addition of photon may induce nonclassicality. Keeping this in mind, we probe whether the same happens for superpostion of coherent states too, specifically, in regard to higher-order nonclassical features that are not observed for them in the previous sections. For example, in this paper,  we have neither observed HOS of Hillery type nor observed HOS of  Hong-Mandel type for the shifted symmetric cat state studied here. We explore the effect of photon addition and addition followed by subtraction on this state and observed that addition of photon  as well as addition followed by subtraction may lead to HOS of Hillery type (cf.  Figs. \ref{fig.HOS-discrete} a-d  and \ref{fig.photon-added}). Here, the effect of photon addition and addition followed by subtraction is studied using a particular witness of higher-order nonclassicality, but it would be straight forward to perform similar studies with respect to the other witnesses of higher-order nonclassicalities.  Now, we may emphasize on the fact that in Fig. \ref{fig.HOS-discrete} a, HOS of Hillery type is observed in the shifted symmetric cat state after the addition of
a single photon and in Fig. \ref{fig.HOS-discrete} b, it is shown
that HOS is observed after addition of the two photons and then subtracting
a single photon for the same state. From Fig.
\ref{fig.HOS-discrete} c it is clear that HOS in photon added then subtracted shifted symmetric cat state can be observed only for a range of values of $\phi$. Similar result is illustrated in Fig. \ref{fig.HOS-discrete} d for the addition of the two photon
and then subtraction of a single photon to shifted symmetric cat state. Fig. \ref{fig.HOS-discrete} c and d further illustrates that
HOS of Hillery type is observed only for symmetric cat state or even coherent state
($\phi=0$), but it is not observed for Yurke-Stoler coherent state
($\phi=\frac{\pi}{2}$) and asymmetric cat state or odd coherent
state ($\phi=\pi$) for $\alpha=0.5$. But higher values of $\alpha$, HOS of Hillery type is observed for the  Yurke-Stoler coherent state  (cf. Fig. \ref{fig.HOS-discrete} a-b). Similar result is obtained for the  asymmetric cat state or odd coherent
state but the result is not depicted here.   On the other hand, in Fig.  \ref{fig.photon-added} a,
effect of addition of increasing number of photons on HOS is systematically illustrated, and in Fig.  \ref{fig.photon-added} b, the effect of $m$-photon addition
followed by two photon subtraction on HOS of the shifted symmetric cat state is illustrated. Both of the cases the depth of HOS witnessis is increasing with photon addition. For the sake of completeness of the study, the effect of photon addition on the HOS of the one-dimensional
continuous superposition of coherent states is also investigated here and
the corresponding result is depicted in the Fig. \ref{HOA-contitnuous} c. 

\section{Conclusion\label{sec:CONCLUSION}}

A coherent state is a classical state, but it's observed that both
the discrete and continuous superposition of coherent states studied
here can show various higher-order nonclassical properties. This observation,
naturally leads to a question- What attributes these nonclassical
properties to a superposition state which can be viewed as superposition
of classical (coherent) states. Clearly, it's nothing but quantum
interference. From earlier, studies \cite{buvzek1995quantum}, we
knew that quantum interference between classical states may lead to
nonclassical state. This fact is further generalized here by demonstrating
that the quantum interference can also lead to higher-order nonclassical
properties. However, which particular feature of higher-order nonclassicality
will be seen in a particular superposition state that would depend
on the nature of superposition. For example, in the shifted symmetric
cat state, we have observed the existence of HOA, but that feature
is not observed here in the one-dimensional continuous superposition of coherent states. It's also observed that HOA and HOSPS are two independent phenomena and existence of one of them does not ensure the existence of the other one. The same is true for HOS, but that seems to be obvious. Further, it is observed that conventional non-Gaussianity inducing operations can induce and/or control higher-order nonclassical properties of the states studied here. Specifically, we may note that HOS is not observed for the discrete superposition of coherent states but it is observed after addition and addition followed by subtraction of the photons.  In brief, various types of higher-order nonclassical features have been observed and their variation with experimentally controllable parameters have been studied for a set of superposition of coherent states that are experimentally realizable. This observation allows us to conclude this article with a hope that the work reported here will be realized experimentally in the near future and the allowed control over the depth of nonclassicality witnesses will be of use in some applications, specially in the context of quantum information processing where the cat states and their variants are used frequently.  

\textbf{Acknowledgment:} A.P. and N.A. thank the Department of Science
and Technology (DST), India, for support provided through the DST
project No. EMR/2015/000393.  A.P. also thanks K. Thapliyal  for some useful technical discussions. K.M. thanks Amit Verma for his help and interest in this work. 

\bibliographystyle{unsrt}
\bibliography{Qcs-final}

\end{document}